\documentclass[aps, prl, superscriptaddress, citeautoscript, 10pt, twocolumn]{revtex4-2}

\usepackage{graphicx}
\usepackage{color}
\usepackage{amsmath}
\usepackage[colorlinks=true, allcolors=blue]{hyperref}

\begin{document}

\title{Polarization faticons:\\
       Chiral localized structures in self-defocusing Kerr~resonators}

\author{Erwan Lucas}
\affiliation{Laboratoire ICB, UMR 6303 CNRS-Université de Bourgogne, 21078 Dijon, France}

\author{Gang Xu}
\author{Pengxiang Wang}
\affiliation{School of Optical and Electronic Information, Huazhong University of Science and Technology, Wuhan, 430074, China}

\author{Gian-Luca Oppo}
\affiliation{SUPA and Department of Physics, University of Strathclyde, Glasgow G4 0NG, Scotland, UK}

\author{Lewis Hill}
\author{Pascal Del'Haye}
\affiliation{Max Planck Institute for the Science of Light, Staudtstraße 2, 91058 Erlangen, Germany}

\author{Bertrand~Kibler}
\affiliation{Laboratoire ICB, UMR 6303 CNRS-Université de Bourgogne, 21078 Dijon, France}

\author{Yiqing Xu}
\author{Stuart G. Murdoch}
\author{Miro Erkintalo}
\author{St\'ephane Coen}
\affiliation{Physics Department, The University of Auckland, Private Bag 92019, Auckland 1142, New Zealand}
\affiliation{The Dodd-Walls Centre for Photonic and Quantum Technologies, Dunedin, New Zealand}

\author{Julien Fatome}
\email{julien.fatome@u-bourgogne.fr}
\affiliation{Laboratoire ICB, UMR 6303 CNRS-Université de Bourgogne, 21078 Dijon, France}

\begin{abstract}
    \noindent We report on numerical predictions and experimental observations of a novel type of temporal localized dissipative structures that manifest themselves in the self-defocusing regime of driven nonlinear optical resonators with two polarization modes. These chiral dissipative solitons, which we term polarization faticons, break both temporal and polarization symmetry and consist of two bright lobes of opposite polarization handedness, interlocked by a domain wall. Our study reveals that faticons are connected to a vectorial modulational instability, from which they can be excited through a collapsing dynamic. Faticons could offer a novel pathway for frequency comb generation in normal dispersion resonators. More generally, they offer new fundamental insights into vectorial localized dissipative structures and could be relevant to other multi-component dissipative systems.
\end{abstract}

\maketitle

\paragraph*{Introduction.}

Self-organized localized dissipative structures (LDSs) are ubiquitous in Nature \cite{akhmediev_dissipative_2008, purwins_dissipative_2010}. They occur in extended systems driven far from thermodynamic equilibrium and are of common occurrence in such diverse fields as physiology \cite{hodgkin_quantitative_1952}, chemistry \cite{lee_pattern_1993}, fluid dynamics \cite{wu_observation_1984, lioubashevski_dissipative_1996}, plasma physics \cite{astrov_plasma_2001}, and ecology \cite{fernandez-oto_strong_2014}. LDSs are especially relevant in nonlinear optics both in single mirror feedback systems \cite{schapers_interaction_2000, ackemann_chapter_2009} and optical cavities \cite{taranenko_spatial_1997, barland_cavity_2002, leo_temporal_2010, herr_temporal_2014, Oppo_PTA_2024} with frequency comb generation \cite{coen_modeling_2013, pasquazi_micro-combs_2018, kippenberg_dissipative_2018}, fiber lasers \cite{grelu_dissipative_2012}, and optical parametric oscillators \cite{bruch_pockels_2021, roy_temporal_2022} as key applications.

Research in optics have further advanced the field by enabling the experimental investigation of multi-component --- or vectorial --- LDSs, whose extra degrees of freedom lead to richer behaviors including anti-phase dynamics \cite{williams_fast_1997, marconi_vectorial_2015}, phase locking and synchronization \cite{cundiff_observation_1999, wright_nonlinear_2022, bao_orthogonally_2019}, or self switching \cite{tsatourian_polarisation_2013, krupa_vector_2017, xu_spontaneous_2021, xu_breathing_2022}. Vectorial LDSs can also emerge through a spontaneous symmetry breaking (SSB) between their components, thereby linking self-localization dynamics to another ubiquitous phenomenology. This connection of LDSs with SSB has only been considered, however, in a handful of theoretical studies \cite{sigler_solitary_2005, skarka_formation_2014, descalzi_breaking_2020, hill_symmetry_2024} and in a single experimental configuration \cite{xu_spontaneous_2021, xu_breathing_2022}.

Here, exploiting optical polarization in a Kerr resonator with normal group-velocity dispersion, we demonstrate theoretically and experimentally the existence of a novel family of SSB-mediated vectorial LDSs, which we refer to as polarization faticons. Remarkably, despite consisting of two bright lobes, with opposite handedness, faticons appear in the self-defocusing regime; their robustness stems from the presence of a domain wall interconnecting the two lobes \cite{garbin_dissipative_2021}. In this way, faticons simultaneously break both temporal and polarization symmetries. On top of providing the first experimental observation of these chiral LDSs, we highlight how they can be generated out of the collapse of a vectorial modulational instability (MI) pattern \cite{hansson_modulational_2018, fatome_polarization_2020}. Finally, we note that polarization faticons could open new avenues for frequency comb generation in the normal dispersion regime \cite{xue_mode-locked_2015, parra-rivas_dark_2016, fulop_high-order_2018, lihachev_platicon_2022} and more fundamentally be relevant in other multi-component systems such as cold gases \cite{goldman_floquet-engineered_2023} or spinors \cite{dogra_dissipation-induced_2019}.

\paragraph*{Model.}

\begin{figure*}
    \centering
    \includegraphics[width=17 cm]{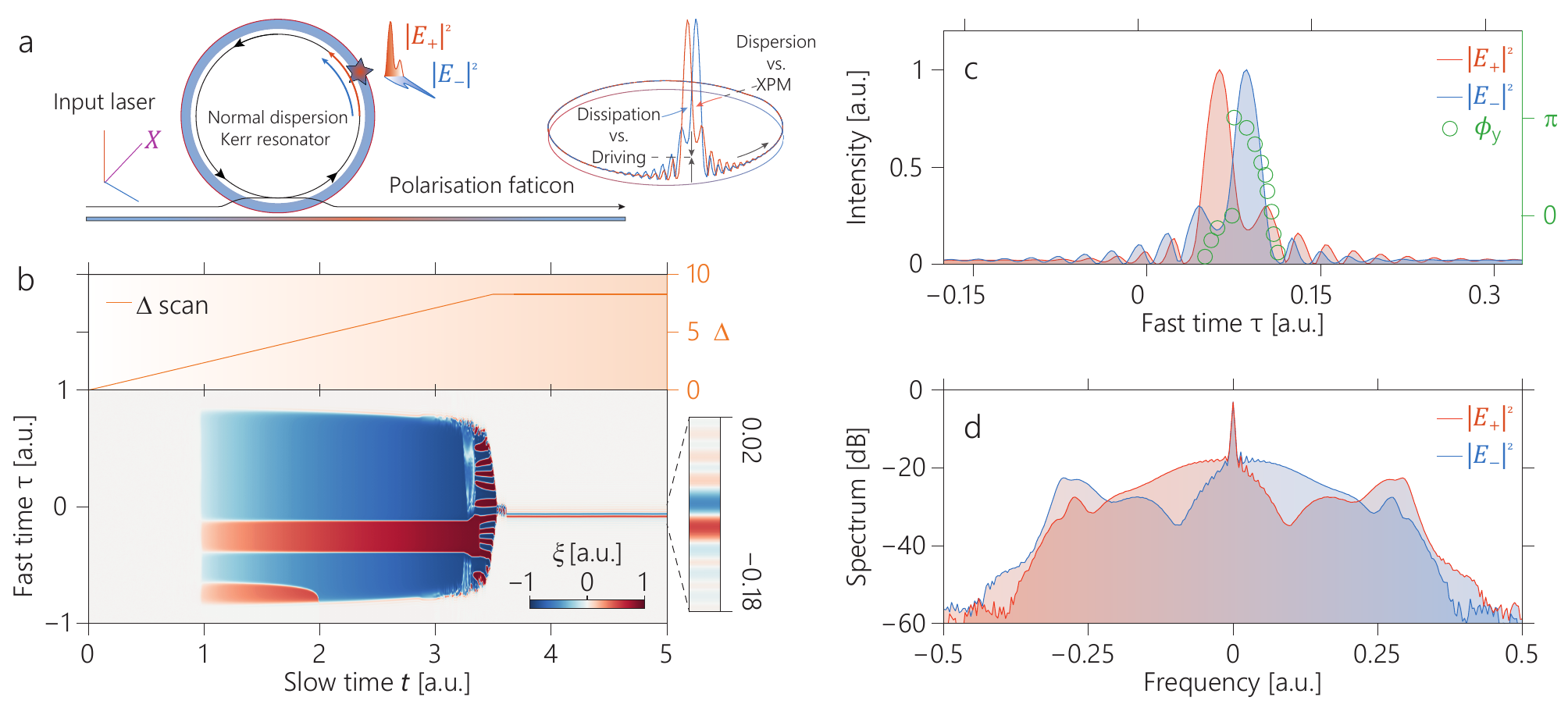}%
    \caption{\textbf{Principle and simulations.} (a) Schematic of a coherently-driven, normally dispersive, two-polarization mode Kerr resonator supporting a faticon. The faticon is a bright localized structure breaking both polarization and temporal symmetries. (b) Numerically calculated slow-time evolution of the fast-time distribution of $\xi=|E_+|^2 - |E_-|^2$ (bottom) as the detuning is ramped then held as shown at the top (pulsed driving, $X=15$, $B=1.85$). (c) Circular polarization modal intensity profiles of the faticon at the end of the simulation shown in (b) ($\Delta=8.3$). The green circles highlight the central $\pi$ phase jump singularity in the corresponding $y$~linear polarization component. (d) Optical intensity spectra corresponding to (c).}
    \label{fig:principle}
\end{figure*}

We consider a passive coherently-driven ring resonator with defocusing Kerr nonlinearity (normal group-velocity dispersion) supporting two circular polarization modes of opposite handedness [see Fig.~\ref{fig:principle}(a)]. The temporal evolution of the complex amplitudes $E_\pm$ of these two modes is described by normalized coupled mean field Lugiato-Lefever equations (LLEs) \cite{haelterman_polarization_1994, geddes_polarisation_1994, garbin_dissipative_2021},
\begin{multline}
    \dfrac{\partial E_{\pm}(t,\tau)}{\partial t} = 
        \biggl[ -i \dfrac{\partial^2}{\partial \tau^2}
               +i \left( |E_{\pm}|^2+B |E_{\mp}|^2 \right)   \\
	-1 -i\Delta \biggr] E_{\pm} + \sqrt{\frac{X}{2}}\,.
    \label{eq:LLEs}
\end{multline}
Here $t$ is a slow time that describes the field's evolution at the scale of the cavity photon lifetime while $\tau$ is a fast time to describe the field's temporal structure at shorter timescales. The terms on the right represent, respectively, group-velocity dispersion, the Kerr nonlinearity, with $B$ the cross-phase modulation (XPM) coefficient \cite{agrawal_nonlinear_2013}, losses, detuning ($\Delta$), and driving, with total driving power~$X$. Note that we assume identical detuning and driving strength for the two modes.

The above equations are well known to exhibit spontaneous symmetry breaking (SSB) (provided $B\neq1$) \cite{kaplan_directionally_1982, del_bino_symmetry_2017, cao_experimental_2017, copie_interplay_2019, garbin_asymmetric_2020, hill_effects_2020}. Upon SSB, symmetric steady states for which $E_+ = E_-$, and that express the symmetry of the driving, become unstable in favor of two mirror-like ($E_+ \rightleftharpoons E_-$) asymmetric solutions where $E_+ \neq E_-$. This phenomenon serves as a prerequisite for the generation of polarization faticons.

\paragraph*{Simulations}

\noindent We first illustrate the essential features of faticons with numerical simulations of Eqs.~(\ref{eq:LLEs}). Specifically, we simulate the nonlinear dynamics when gradually ramping then holding the cavity detuning up to a certain value, a procedure typical of microresonator excitation of bright cavity solitons (CSs) and optical frequency combs \cite{herr_temporal_2014, pasquazi_micro-combs_2018, kippenberg_dissipative_2018}. The results are illustrated in Fig.~\ref{fig:principle}(b) for a pulsed drive mimicking the conditions of the experiments that will follow [note that faticons also exist with a continuous-wave (cw) drive]. To highlight the polarization asymmetry --- and the handedness --- of the intracavity field, we plot the difference between the modal intensities, $\xi = |E_+|^2 - |E_-|^2$ (fast time vs slow time), with the corresponding evolution of the detuning at the top.

The field is initially in a symmetric state ($\xi=0$) until polarization SSB occurs for $\Delta \sim 2.4$ (at $t\simeq 1$). At this point, the field abruptly segregates into domains, each realizing one of the two stable asymmetric homogeneous steady state (HSS) solutions of Eqs.~(\ref{eq:LLEs}), with opposite handedness \cite{haelterman_polarization_1994}. These polarization domains are separated by sharp temporal transitions, known as dissipative polarization domain walls (PDWs) \cite{haelterman_pdw_1994, gilles_polarization_2017, garbin_dissipative_2021, fatome_temporal_2023}. As the detuning ramps up, one domain collapses and the polarization contrast increases, but the situation stays qualitatively the same until $\Delta \sim 7.8$ (at $t\simeq 3.3$) where a periodic anti-correlated vectorial pattern resulting from a polarization MI invades the entire circulating pulse~\cite{kockaert_isotropic_1999, fatome_polarization_2020, hansson_modulational_2018}. Shortly after, the pattern collapses into a short localized solitonic structure made up of two lobes of opposite handedness, the polarization faticon, with the rest of the field returning to a low power symmetric state. As the detuning is then held steady at $\Delta=8.3$, the faticon persists in the resonator. 

Figure~\ref{fig:principle}(c) displays the temporal intensity profile of the two modes of the faticon ($|E_+|^2$ and $|E_-|^2$) at the end of the simulation shown in Fig.~\ref{fig:principle}(b). The faticon consists of a bright localized polarization twist made up of two interlocked polarized lobes of opposite handedness flanked by oscillatory tails, and separated by a central PDW. The structure is infrangible: one lobe cannot exist without the other. It has a chiral nature, exhibiting simultaneously broken polarization and time symmetries, with mirror structures equally possible. This is in contrast with the bright structures reported in~\cite{xu_spontaneous_2021} with anomalous dispersion where only the polarization symmetry is broken. We note that, by symmetry, the two circular modal amplitudes are equal ($E_+=E_-)$ at the center of the faticon, where the field locally exhibits a linear polarization state (say along~$x$). We have evidenced this feature by plotting the phase of the corresponding orthogonal ($y$) linear polarization component [green circles in Fig.~\ref{fig:principle}(c)], revealing a topological $\pi$ phase jump at that location.

The symmetries of the faticon structure can be further appreciated in the frequency domain, with Fig.~\ref{fig:principle}(d) showing the intensity spectra of the two circular components. The leading and trailing lobes, where either $|E_+|^2$ or $|E_-|^2$ dominate, are seen to be respectively red- and blue-shifted symmetrically. These frequency shifts are caused by the XPM interaction between the two lobes. Additionally, strong dispersive components appear at the edges of each spectrum, typical of shock front dynamics in the defocusing regime~\cite{malaguti_dispersive_shock_2014, xu_frequency_2021, bunel_broadband_2024}, and corresponding to the oscillating tails of the faticon.

\begin{figure}[t]
    \centering%
    \includegraphics[width=\linewidth]{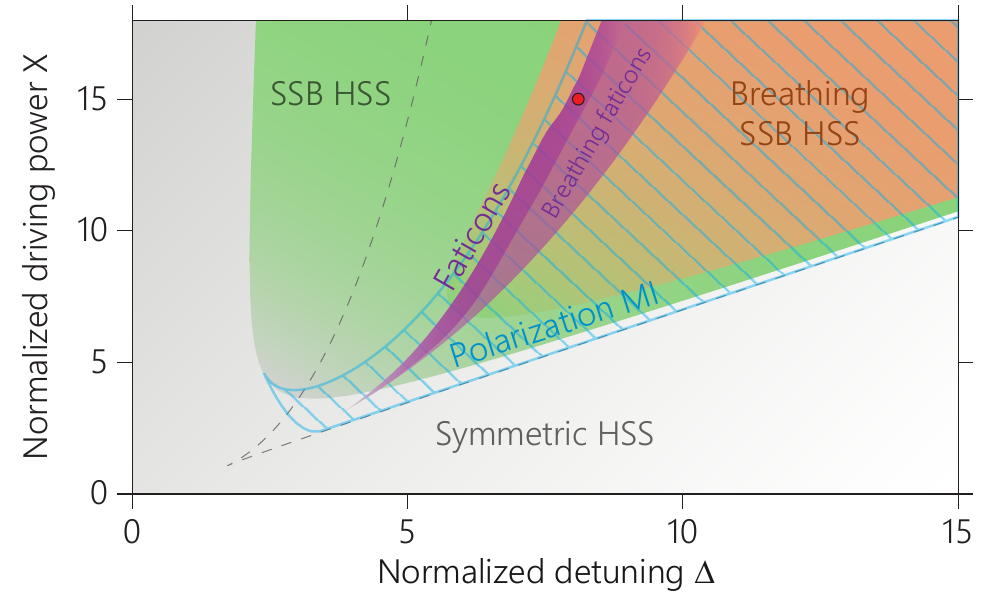}
    \caption{\textbf{Theoretical phase diagram of Eqs.~(\ref{eq:LLEs}).} Dark purple: faticons (the circle corresponds to our experimental conditions). Light purple: Breathing faticons. Gray area: symmetric HSSs (also present and stable between the gray dashed lines). Green: asymmetric HSSs. Peach: Oscillatory HSSs. The blue stripes indicate where the underlying HSS solutions are modulationally unstable. $B=1.85$.}
    \label{fig:existence_map}
\end{figure}

To gain a deeper understanding, we have systematically sought stable faticon solutions of Eqs.~(\ref{eq:LLEs}) with a Newton solver \cite{mcsloy_computationally_2002} for a range of detunings~$\Delta$ and cw driving powers~$X$. From these numerical results, we have plotted the faticons' existence range in Fig.~\ref{fig:existence_map} against that of different HSSs. Stable and breathing faticons exist in a relatively narrow band of parameter values (dark and light purple areas, respectively) mostly overlapping with the middle of the region of parameter space where symmetry-broken HSSs arise (green) and close to the limit where (vectorial) polarization MI emerges and destabilize the HSSs (stripes). Also shown is the region where asymmetric HSSs develop a breathing instability (peach). In the gray-shaded area, only symmetric HSSs are found.
Finally, in between the gray dashed lines, the presented multi-valued upper states co-exist with a symmetric lower state, which is always stable [see End Matter Fig.~\ref{fig:NL_resonance} for a more detailed cross section of the diagram].
Together with the results of Fig.~\ref{fig:principle}(b), this suggests that faticons, like bright scalar temporal CSs, correspond to a homoclinic connection between the symmetric HSS and a single period of a (vectorial) MI pattern.

\paragraph*{Experiment}

\begin{figure}[t]
    \centering
    \includegraphics[width=\linewidth]{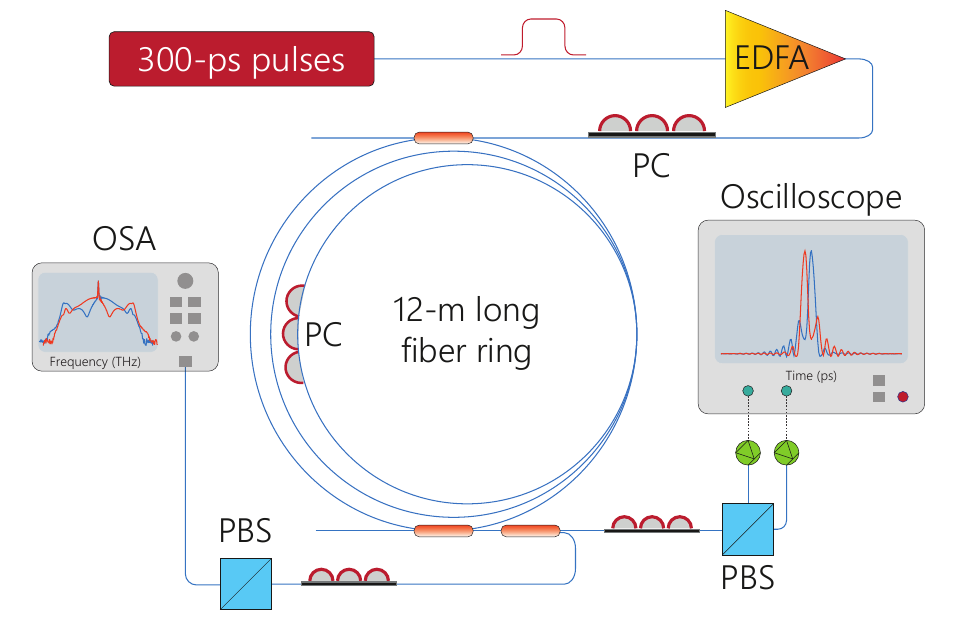}
    \caption{\textbf{Experimental setup.} Schematic illustration of the experimental implementation. EDFA: erbium-doped fiber amplifier, PC: polarization controller, PBS: polarizing beam splitter, OSA: optical spectrum analyzer.}
    \label{fig:setup}
\end{figure}

Figure~\ref{fig:setup} shows a simplified diagram of our experimental setup. Our resonator consists of 12-m of single-mode spun optical fiber exhibiting normal group-velocity dispersion at the $1552.4$~nm driving wavelength. The XPM coefficient $B$ was estimated at $1.85$ by comparing experiments with theory. The resonator includes a 90/10 coupler for injection of the driving field and a 1\,\% tap coupler to monitor the intracavity dynamics, yielding an overall round-trip time of 57~ns and a finesse of~27. It is synchronously driven by 300-ps flat top pulses, amplified to a peak driving power of~$4.9$~W, corresponding to $X=15$ (see End Matter for more detail). The setup also includes three polarization controllers (PCs). The one placed inside the resonator is configured to implement a $\pi$ phase-shift linear birefringent defect to operate in the symmetry protected regime described in \cite{coen_nonlinear_2024}. With that defect, the intracavity field's handedness swaps at each round trip, resulting in the averaging of all asymmetries and enabling the realization of polarization SSB in ideal conditions, i.e., equal driving and detuning for both modes as assumed in Eqs.~(\ref{eq:LLEs}). (Note that in the results presented below this swapping dynamics is unwrapped for clarity purpose.) Two other PCs are placed at the input and output of the resonator, respectively. At the input, we align the polarization of the driving field so as to equally excite both circular modes of the resonator, while at the output, together with polarizing beam splitters (PBSs), we can split the output field into either its circular or linear polarization components that can then be detected with fast photodiodes or sent to an optical spectrum analyzer.

\paragraph*{Experimental results.}

\begin{figure}[t]
    \centering%
    \includegraphics[width=\linewidth]{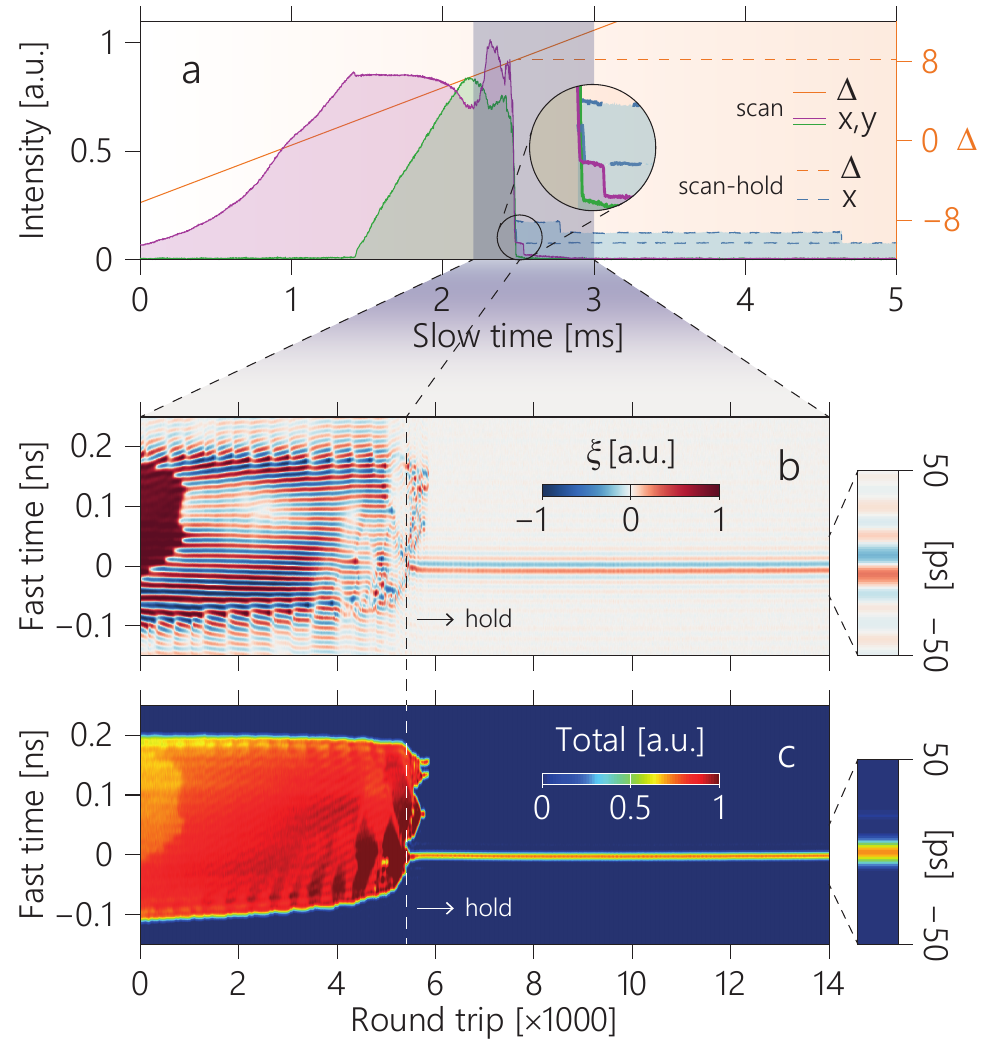}

    \caption{\textbf{Experimental excitation of a polarization faticon.} (a) Average output power of the $x$ and $y$ polarization components (left axis) and detuning (right axis) for a continuous scan of detuning across a resonance (solid curves) and for a scan-hold sequence (dashed curves). (b), (c) Detail of the scan-hold dynamics [corresponding to the gray zone in (a)] in terms of polarization contrast (fast-time vs round-trip number) and total intensity. A faticon emerges around round trip 5,500, from where the detuning is held constant (dashed vertical line) at $\Delta \approx 8.1$.}

    \label{fig:scan_exp}
\end{figure}

To identify the range of detunings where faticons arise experimentally, we first performed a full scan of a cavity resonance. This is illustrated by the solid set of curves in Fig.~\ref{fig:scan_exp}(a), representing the measured evolution of the average intracavity power resolved in terms of the $x$ (purple) and $y$ (green) linear polarization components over time, as the detuning~$\Delta$ is scanned (orange, right axis). Initially, only the $x$ polarization is excited, compatible with the balanced driving of the circular modes. $1.4$~ms into the scan ($\Delta \approx 2.6$), we observe a threshold-like behavior for the $y$ component, which is associated with the occurrence of SSB (see~\cite{coen_nonlinear_2024}). At $2.3$~ms, the signal becomes noisier, signaling polarization MI. Shortly after ($2.5$~ms), the output power suddenly collapses, immediately followed by a very short step (see purple curve in the zoomed inset) --- a characteristic feature often associated with the emergence of LDSs in nonlinear passive resonators \cite{herr_temporal_2014}.

We then repeated those measurements with a scan-hold detuning sequence (see End Matter for implementation) as in the simulation of Fig.~\ref{fig:principle}(b), with the detuning held steady after the collapse. These results are plotted as dashed curves in Fig.~\ref{fig:scan_exp}(a) (the $y$ component is omitted for clarity). Two realizations are shown (light blue curves), one of which revealing multiple discrete steps associated with the simultaneous excitation of several faticons (see End Matter). For the single step case, more details of the experimental field dynamics around the collapse are revealed in Fig.~\ref{fig:scan_exp}(b), where we plot the polarization handedness contrast~$\xi=|E_+|^2 - |E_-|^2$, temporally resolved across the intracavity pulse, for the section of the scan-hold detuning sequence highlighted in gray in panel~(a). On the far left of the plot, the near homogeneous patch reveals that part of the field expresses an asymmetric HSS, before an alternating handedness pattern characteristics of polarization MI invades the whole pulse. The pattern persists as the detuning keeps increasing until roundtrip 5,500 where it collapses into what can be seen as a single period of the pattern, with two lobes of opposite handedness --- the polarization faticon --- with the rest of the field returning to a symmetric state ($\xi=0$). As the detuning is then held steady, the faticon persists in the resonator for thousands of round trips [constant blue level in panel~(a)], thus demonstrating its self-localized nature. For completeness, we show in Fig.~\ref{fig:scan_exp}(c) the corresponding evolution of the total intensity $|E_+|^2 + |E_-|^2$. Remarkably, while the collapse event is clearly visible, there is no discernible evidence of any patterned state, further confirming the vectorial (polarization) character of the faticons' dynamics.

\begin{figure}[t]
    \centering%
    \includegraphics[width=\linewidth]{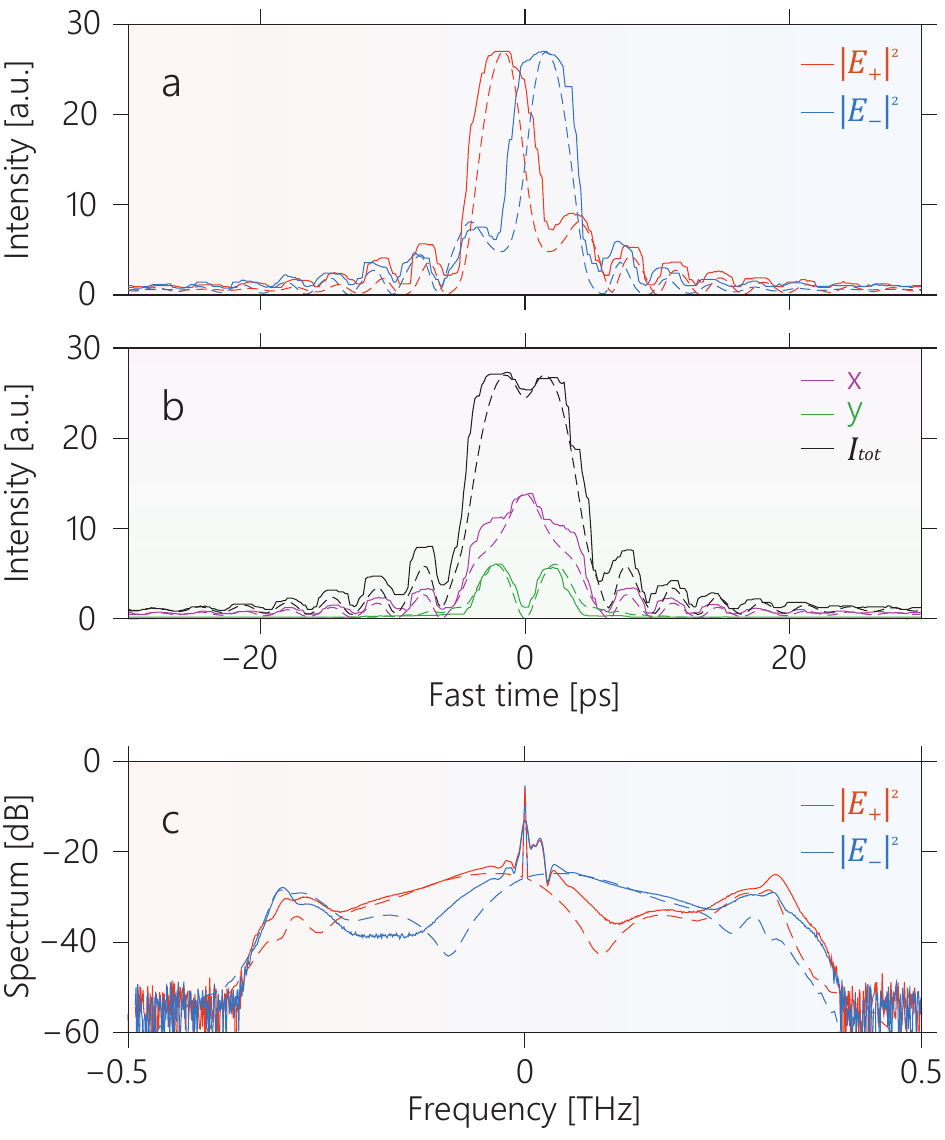}%
    \caption{\textbf{Experimental characterization of the polarization faticon.} (a), (b) Temporal intensity profiles of the faticon in terms of the circular and linear polarization components, respectively. In (b), the black curve also shows total intensity. (c) Optical intensity spectra of the circular components. In all panels, measurements (solid) are compared to simulations (dashed).}
    \label{fig:spectre}
\end{figure}

The experimentally measured dynamics observed in Fig.~\ref{fig:scan_exp} show excellent qualitative agreement with the simulations presented in Fig.~\ref{fig:principle}. For more insights, Fig.~\ref{fig:spectre} shows a detailed spectro-temporal characterization of the faticon excited in our experiments. The solid curves in panels~(a) and (b) show high resolution measurements of the temporal intensity profiles of, respectively, the circular and linear polarization components of the faticon [with panel~(b) also showing total intensity]. These have been obtained with a 500-GHz bandwidth optical sampling oscilloscope and agree remarkably well with numerical predictions (dashed curves). The experimental traces reveal two interleaved lobes of opposite handedness, each with a temporal duration of about 5~ps, and with leading and trailing oscillations. We note that the mirror structure, where the two circular components are swapped, is observed equally likely in the experiment, depending on random initial fluctuations. In Fig.~\ref{fig:spectre}(b), we also note a central zero for the $y$ component, indicating the equality of the field's circular components at the center of the faticon, again a telltale sign of the symmetry and topology of the structure we are observing. Finally, Fig.~\ref{fig:spectre}(c) shows the experimental optical intensity spectra of the circular components of the faticon, again compared with numerical modeling. The $+$ mode, which is dominant in the leading lobe of the faticon, is preferentially red-shifted, and vice-versa. Strong dispersive radiations are also present at the edges of the spectra, associated with the tail oscillations visible in the temporal traces. Overall, the agreement with theoretical expectations is excellent (see Fig.~\ref{fig:principle}).

\paragraph*{Conclusion.}

Using a coherently-driven Kerr resonator with two polarization modes, we have presented the first experimental evidence of faticons, a new type of vectorial LDSs arising through SSB in the defocusing regime. In our implementation, faticons consist of two interlocked polarized lobes of opposite handedness, separated by a central domain wall. Notably, faticons have a chiral nature, and simultaneously break both polarization (modal) and temporal symmetries. The temporal symmetry breaking and cross-phase modulation across the domain wall \cite{garbin_dissipative_2021} enables the structure to be self-localized in the defocusing regime. Together with the symmetry-broken CSs reported in \cite{xu_spontaneous_2021, xu_breathing_2022} in the focusing regime, they form the only examples of experimental observations of SSB of LDSs in two-component physical systems. Our observations are in remarkable agreement with numerical simulations based on incoherently-coupled, two-components LLEs. In particular, we have highlighted how faticons emerge as a single period of a vectorial MI periodic pattern in a way akin to scalar CSs. By symmetry, faticons come in two different mirror configurations and could be advantageous for dual comb generation in spectral windows with normal chromatic dispersion. On a more fundamental level, and given the LLE's ubiquitous character, faticons could have relevance to other bimodal systems including two-component atomic gases or spinors \cite{goldman_floquet-engineered_2023, dogra_dissipation-induced_2019}. Finally, while beyond the scope of the current study, simulations reveal the existence of a larger family of faticon-like structures, with more than two lobes, in a collapsed snaking arrangement, as well as breathing states. In this regard, our work paves the way for future studies of the dynamics of this novel family of vectorial dissipative structures.

\begin{acknowledgments}
    EL, BK, and JF acknowledge financial support from the CNRS, International Research Project WALL-IN, the Conseil Régional de Bourgogne Franche-Comté (mobility program) and the FEDER-FSE Bourgogne 2014/2020 programs. LH and JF acknowledge the Max Planck Institute and the CNRS (Salto exchange program). We also acknowledge financial support from The Royal Society of New Zealand, in the form of Marsden Funding (18-UOA-310 and 23-UOA-053).
\end{acknowledgments}

\bibliography{cleanrefs}

\clearpage

\appendix

\section*{End Matter}

\setcounter{figure}{0}
\renewcommand\thefigure{S\arabic{figure}}

\begin{figure}[h]
    \centering%
    \includegraphics[width=\linewidth]{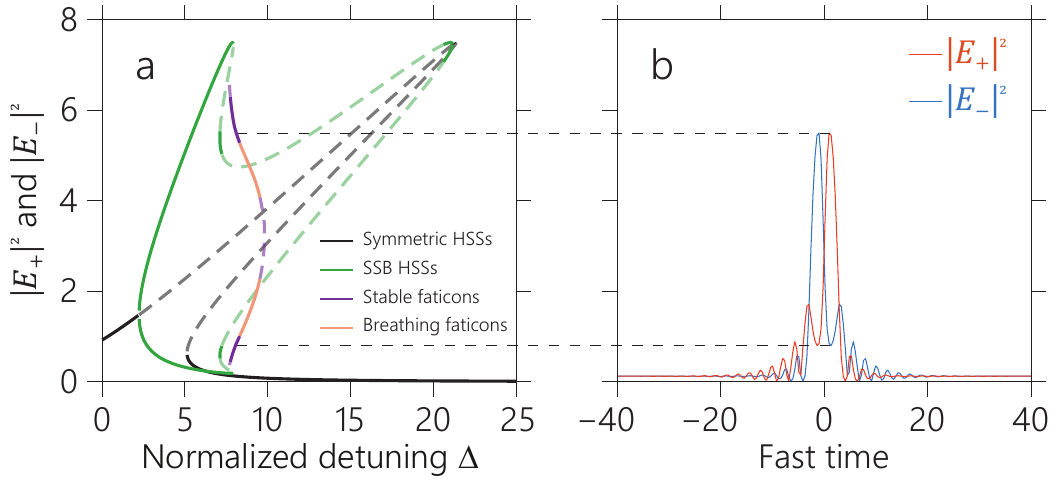}
    \caption{\textbf{Bifurcation diagram of Eqs.~(\ref{eq:LLEs}) for $X=15$ and $B=1.85$.} (a) Intensity of the circular polarization components $|E_+|^2$ and~$|E_-|^2$ versus detuning~$\Delta$ for the symmetric (black curve) and asymmetric (green) HSSs as well as for faticon solutions (purple and peach). Dashed lines indicate unstable solutions. (b)~shows the temporal intensity profile of the faticon corresponding to $\Delta=8.1$.}
    \label{fig:NL_resonance}
\end{figure}

\noindent \textbf{Bifurcation diagram.} To illustrate in more detail how faticons are organized with respect to HSSs, we report in Fig.~\ref{fig:NL_resonance}(a) a bifurcation diagram that corresponds to a cross section of Fig.~\ref{fig:existence_map} (for cw driving power~$X=15$). The plot shows, as a function of detuning~$\Delta$, the intensity of the circular polarization components $|E_+|^2$ and~$|E_-|^2$ for the symmetric (black) and asymmetric HSSs (green) as well as for the stable and breathing faticons (purple and peach, respectively). For the faticon, the top and bottom part of the curve correspond, respectively, to the peak and trough of the main lobe [see Fig.~\ref{fig:NL_resonance}(b)]. Dashed lines in panel~(a) indicate unstable solutions (for HSSs, instability arise mostly through polarization MI). The bifurcation diagram shows that SSB of the HSS occurs around $\Delta \approx 2.3$. Stable faticons exist for detunings $7.7\le\Delta\le8.3$ around the region where asymmetric HSSs exhibit multistability. Note that the limited range of existence of stable faticons highlight some of the technical challenges of our experimental work.

\medskip

\noindent \textbf{Fiber parameters.} Our resonator is made up of nearly-isotropic spun optical fibre. At the driving wavelength of $1552.4$~nm, the fiber presents a group-velocity dispersion coefficient of $\beta_2 \approx 46\ \mathrm{ps^2/km}$ and a nonlinear Kerr coefficient $\gamma \approx 4\ \mathrm{W^{-1}\,km^{-1}}$. 

\medskip

\noindent \textbf{Driving beam.} For driving, we use a cw distributed feedback fiber laser (Koheras) at $1552.4$~nm wavelength. Its $<1$~kHz linewidth is much narrower than the 640~kHz resonance linewidth of the fiber resonator, ensuring coherent driving. To reach higher peak driving power, we carve 300-ps flat top pulses into the cw beam using a Mach–Zehnder amplitude modulator, driven by a pulse pattern generator (PPG) synchronized with the cavity round-trip time. For the measurements of Fig.~\ref{fig:spectre}, an additional 10-GHz weak sinusoidal phase modulation is imparted into the driving pulses to create temporal trapping sites for the faticons, which mitigates minor residual desynchronization between the driving pulses and the fiber resonator \cite{erkintalo_phase_2022}. Before injection, we also amplify the driving pulse train with an erbium-doped fiber amplifier (EDFA) and an additional spectral filter removes amplified spontaneous emission noise. 

\medskip

\noindent \textbf{Detuning control.} To control and stabilize the detuning parameter~$\Delta$ in our experiments, we proceed in two steps. First, the wavelength of the driving laser is actively locked to a linear resonance of the fiber resonator using the Pound-Drever-Hall (PDH) technique. This is based on a small fraction of the cw driving laser that counter-propagates in the resonator with respect to the main (nonlinear) beam. Second, the main driving beam is frequency shifted using single-sideband modulation before generating the flat-top pulses. By adjusting the modulation frequency (around 30~GHz in our case) we can reach arbitrary detuning values. Scan-hold detuning sequences are implemented by programming sweeps of the modulation frequency with an arbitrary signal generator.

\begin{figure}[b]
    \centering
    \includegraphics[width=\linewidth]{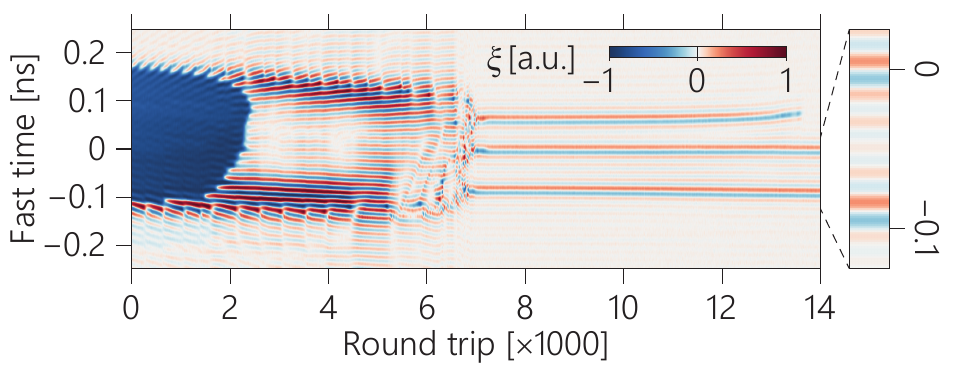}
    \caption{\textbf{Experimental generation of multiple faticons.} Polarization handedness contrast~$\xi$ (fast-time vs round-trip number) showing the simultaneous generation of three faticons using the same parameters and scan-hold detuning sequence as in Fig.~\ref{fig:scan_exp}.}
    \label{fig:molecule}
\end{figure}

\medskip
 
\noindent \textbf{Generation of multiple faticons.} As the polarization faticons emerge randomly from a polarization MI pattern when the detuning is scanned, it is possible for several faticons to be generated simultaneously. Figure~\ref{fig:molecule} presents an experimental observation of such an occurrence using the same type of polarization handedness ($\xi$) plot (fast-time versus slow-time) as in Fig.~\ref{fig:scan_exp}(b). The  experimental conditions and the scan-hold detuning sequence were also identical. As in Fig.~\ref{fig:scan_exp}(b), we initially observe an asymmetric HSS (near uniform patch on the left side of the diagram), which turns into a pattern upon ramping of the detuning. Subsequently, as the pattern collapses, we can observe the simultaneous emergence of three faticons. This occurs around round trip~7000, when $\Delta \approx 8.1$. The detuning is help steady from then on. One faticon decays at round trip~13,500 but two persists further within the resonator in a stable configuration, leading to the formation of what could be characterized as a faticon molecule.

\end{document}